\documentclass[%
 reprint,
 amsmath,amssymb,
 aps,
]{revtex4-2}

\usepackage{graphicx}
\usepackage{dcolumn}
\usepackage{bm}

\usepackage{lineno,hyperref}
\usepackage{braket}
\usepackage{amsmath,amssymb}
\usepackage{color}
\usepackage{physics}
\usepackage[T1]{fontenc}

\begin{document}

\preprint{APS/123-QED}

\title{Bose-Einstein Condensation of Europium}

\author{Yuki Miyazawa$^{1}$}
\author{Ryotaro Inoue$^{1}$}
\author{Hiroki Matsui$^{2}$}
\author{Gyohei Nomura$^{2}$}
\author{Mikio Kozuma$^{1,2}$}

\affiliation{%
	$^{1}$Institute of Innovative Research, Tokyo Institute of Technology, 4259 Nagatsuta, Midori, Yokohama, Kanagawa 226-8503, Japan}
	
\affiliation{%
	$^{2}$Department of Physics, Tokyo Institute of Technology, 2-12-1 O-okayama, Meguro, Tokyo 152-8550, Japan}


\begin{abstract}
 We report the realization of a Bose-Einstein condensate of europium atoms, which is a strongly dipolar species with unique properties, a highly symmetric $[\mathrm{Xe}]\ 4f^7 6s^2\ {}^8\mathrm{S}_{7/2}$ electronic ground state and a hyperfine structure.
 By means of evaporative cooling in a crossed optical dipole trap, we produced a condensate of ${}^{151}$Eu containing up to $5\times 10^4$ atoms.
 The scattering length of ${}^{151}$Eu was estimated to be $a_s = 110(4)\, a_\mathrm{B}$ by comparing the velocities of expansion of condensates with different orientations of the atomic magnetic moments.
 We observed deformation of the condensate in the vicinity of the Feshbach resonance at 1.32\,G with a width of 10\,mG.
\end{abstract}

\maketitle

The realization of Bose-Einstein condensates (BECs) in highly magnetic atoms has opened new frontiers in quantum simulations \cite{Lahaye2008, B_ttcher_2020, Norcia2021}.
Pioneering work began with chromium (Cr) \cite{Griesmaier2005}, in which a strong dipole-dipole interaction (DDI) was observed as a d-wave symmetric collapse of the condensate \cite{Lahaye2008} originating from the large dipole moment of $\mu = 6\,\mu_\mathrm{B}$.
Bose-Einstein condensates of the magnetic lanthanide atoms dysprosium (Dy) and erbium (Er) were also produced \cite{Lu2011, Aikawa2012}.
Since these atoms possess large magnetic dipole moments ($\mu = 10\,\mu_\mathrm{B}$ (Dy) and $\mu = 7\,\mu_\mathrm{B}$ (Er)) and large atomic masses ($m=160\sim170\,\mathrm{u}$), their dipolar lengths $a_{dd} = m\mu_0 \mu^2 / 12 \pi \hbar^2$ \cite{Lahaye2008} are much larger than that of Cr and comparable to their s-wave scattering lengths.
This strong magnetic property enables investigations of unprecedented quantum phases: self-bound droplets and supersolid crystals \cite{B_ttcher_2020}.
Optical lattice systems with these magnetic atoms are also areas of intense research \cite{doi:10.1126/science.aac9812}.

Dipolar BECs with a spin degree of freedom, namely, spinor dipolar BECs, are the focus of another research direction.
Novel spin physics is expected to emerge in the competition of spin-dependent contact interactions and DDIs \cite{Kawaguchi2006, Kawaguchi2006a, Takahashi2007}.
In general, the ratio of these two interactions cannot be tuned under the experimental conditions used for spinor dipolar BECs; observable phenomena vary depending on the atomic species.
Thus far, spinor dipolar BECs with magnetic atoms have been realized only in Cr \cite{Pasquiou2011, Lepoutre2018a}, whose spin-dependent contact interactions are much larger than DDIs; i.e., its dipolar length of $a_{dd} = 15\,a_\mathrm{B}$ is one order of magnitude smaller than the difference between the scattering lengths $a_{0,2,4,6}$ \cite{Werner2005, Pasquiou2010, Paz2014}.
To gain an in-depth understanding of spinor dipolar BECs, it is essential to realize such phenomena in atoms whose spin-dependent contact interactions are smaller than DDI.

Europium (Eu), a magnetic lanthanide with a dipole moment of $\mu = 7\,\mu_\mathrm{B}$, features a highly symmetric $[\mathrm{Xe}]\ 4f^7 6s^2\ {}^8\mathrm{S}_{7/2}$ electronic ground state.
This unique property distinguishes Eu from other magnetic atoms in the context of spinor dipolar BECs.
Since the half-filled $4f$ shell is screened by the outer closed $6s$ shell, the molecular potentials of Eu associated with different spin channels are almost identical \cite{doi:10.1063/1.3282332, Suleimanov2010}, in contrast to those of Cr \cite{Pavlovifmmodecuteclseci2004}, leading to a smaller spin-dependent interaction.
In addition, the dipolar length of Eu, $a_{dd} = 60\,a_\mathrm{B}$, is four times larger than that of Cr due to a larger dipole moment and greater atomic mass.
Consequently, the formation of DDI-dominant spinor dipolar BECs is strongly suggested in the case of Eu.
We note that the behavior of spinor dipolar BECs of other magnetic lanthanide atoms may differ considerably from those of Cr and Eu due to their orbital anisotropy \cite{Li2018}.

Another feature of Eu is its hyperfine structure ($F=1-6$ for ${}^{151}$Eu and ${}^{153}$Eu) in the ground state, which is useful for the study of spin physics.
The quadratic Zeeman shift induced by hyperfine interactions can be used to prepare specific magnetic sublevels \cite{Baier2018, Patscheider2020}.
Quantum simulations of quantum magnetism \cite{Gabardos2020, Patscheider2020} are expected to be possible with a wide range of spin lengths $F=1-6$.
The hyperfine interaction, in principle, enables control of contact interactions with microwaves even under dc zero magnetic fields \cite{Papoular2010}, which could greatly expand the scope of research on spinor dipolar BECs.
Note that the hyperfine spacings of Eu are relatively small ($\leqq121$\,MHz \cite{Jin2002}), which can mitigate technical difficulties in the above applications.

In this letter, we report the attainment of ${}^{151}$Eu BECs containing $5\times 10^4$ atoms.
In addition, we measure the s-wave scattering length, observe the low-magnetic-field Feshbach spectrum, and demonstrate control of the scattering length with the Feshbach resonance.

Our experimental procedure to create a BEC of Eu consists of a narrow-line magneto-optical trap (MOT), transfer of atoms from the MOT to a single optical dipole trap (ODT), and subsequent evaporative cooling in a crossed ODT, similar to experiments involving other magnetic lanthanide atoms \cite{Aikawa2012, Davletov2020}.
We first loaded atoms into the narrow-line red-MOT \cite{Miyazawa2021} using the $a^8\mathrm{S}_{7/2}\leftrightarrow z^{10}\mathrm{P}_{9/2}$ transition at 687\,nm with a natural linewidth of 97\,kHz, which was operated simultaneously with the yellow-MOT \cite{Inoue2018} and the Zeeman slowing for the optically pumped metastable atoms \cite{SM}.
Our red-MOT contains $7\times10^7$ atoms on the electronic ground state at a temperature of $10\,\mathrm{\mu K}$, where the atoms were spontaneously polarized to the lowest Zeeman substate $\ket{F=6, m_F=-6}$ with the aid of the gravity \cite{Miyazawa2021, Dreon_2017}.

Our atom trap system consisted of horizontal and vertical ODTs, with a light source derived from a single-frequency fiber laser at a wavelength of $1550\, \mathrm{nm}$.
The horizontal ODT was formed by a horizontally propagating focused beam with an initial power of $10\, \mathrm{W}$.
The beam had a waist of 31(25)\,$\mathrm{\mu m}$ along the horizontal (vertical) direction and was horizontally linearly polarized.
The vertical ODT was formed by a beam tilted at $12^\circ$ to the vertical with a maximum power of $1.6\, \mathrm{W}$.
The beam had a waist of $42\,\mathrm{\mu m}$ and was polarized orthogonal to the horizontal ODT beam polarization.
The power of each beam was controlled independently by an acousto-optic modulator.
To avoid the spatial interference between the two beams, their frequencies were shifted by $160\, \mathrm{MHz}$ relative to each other.

The atoms trapped in the MOT were then loaded into the horizontal ODT.
Loading was performed by overlapping the MOT and horizontal ODT for an optimal loading time of $40\, \mathrm{ms}$.
After loading, we turned off the MOT beams and the quadrupole magnetic field, and we applied a homogeneous magnetic field of 3\,G along the vertical direction to preserve the spin polarization of the sample.
We obtained $3.5\times 10^6$ atoms at a temperature of $50\, \mathrm{\mu K}$ in the horizontal ODT.
The trap frequency under loading conditions was measured to be $(\nu_x, \nu_y, \nu_z) = (0.03, 1.5, 1.8)\, \mathrm{kHz}$, where the $x$, $z$, and $y$ axes correspond to the horizontal ODT beam axis, the vertical axis, and the axis perpendicular to them, respectively.
Using these values, we estimated the potential depth of the trap, peak density, and peak phase space density to be $350\,\mathrm{\mu K}$, $3.3\times 10^{13}\, \mathrm{cm^{-3}}$, and $2.7\times 10^{-4}$, respectively.
These were our starting conditions for the following evaporative cooling process.

Evaporative cooling was carried out by gradually reducing the power of the horizontal ODT beam.
In the middle of the evaporative cooling process, we turned on the vertical ODT, which intersected with the horizontal ODT, and atoms were then concentrated in the crossing region.
The overall evaporation sequence took approximately $8\, \mathrm{s}$ \cite{SM}.
After that, we turned off the trapping beams, let the atomic cloud expand, and rotated the magnetic field into the imaging axis ($y$-axis) for absorption imaging.
The atoms were then illuminated by a $\sigma_-$ polarized probe laser beam along the $y$-axis at a wavelength of $460\, \mathrm{nm}$, which was resonant with the $\ket{a^8 \mathrm{S}_{7/2}, F=6, m_F=-6} \leftrightarrow \ket{y^8 \mathrm{P}_{9/2}, F'=7, m_{F'}=-7}$ cyclic transition.
The absorption by the atoms cast a shadow on an imaging device.

\begin{figure}[htb]
	\includegraphics{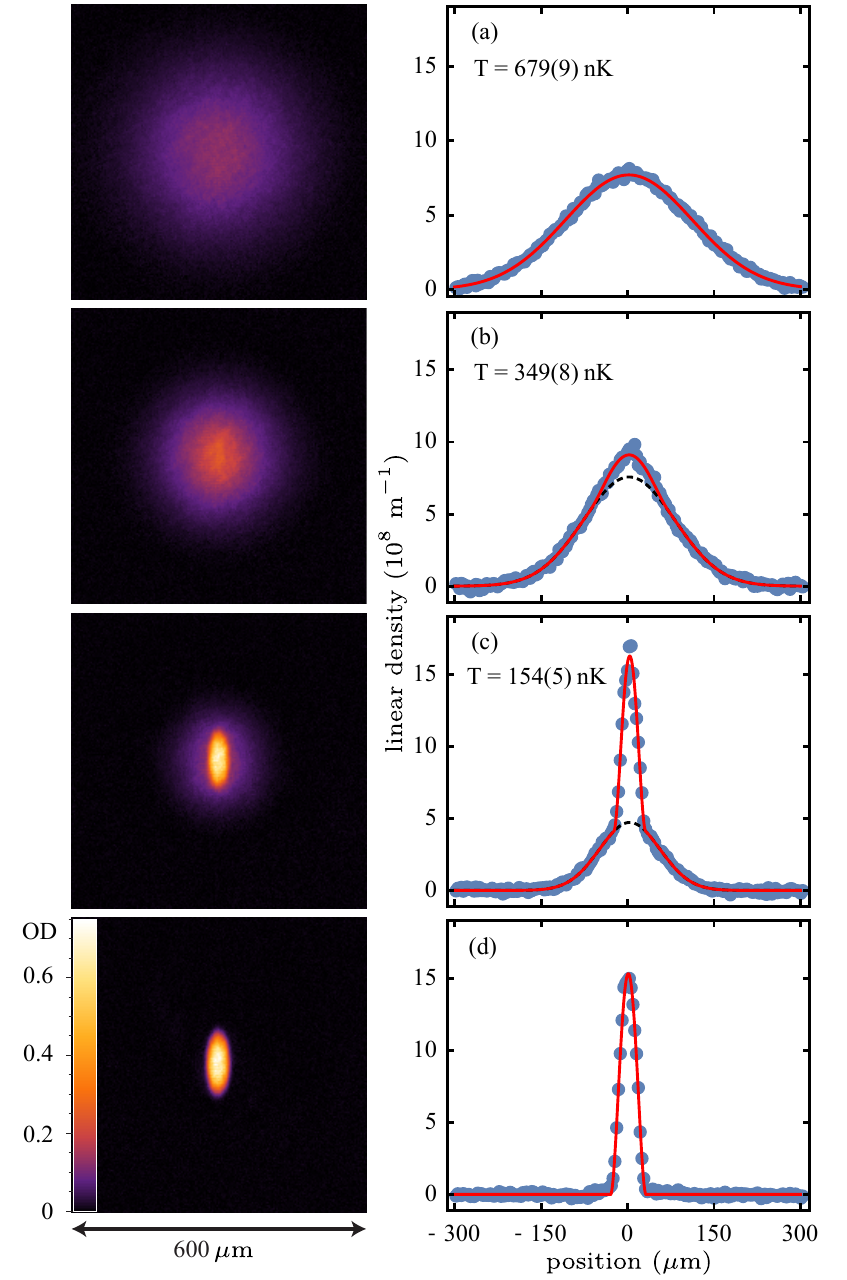}
	\caption{\label{fig:1}
	Absorption images and integrated density profiles showing the BEC phase transition at different temperatures.
	The absorption images are an average of 10 images taken after $18\, \mathrm{ms}$ of expansion.
	The color bar shows the optical density.
	The solid lines are fits of the data using Gaussian (a), bimodal (b) and (c), and Thomas-Fermi (d) distributions.
	The dotted lines represent the Gaussian part of the bimodal fit, describing the thermal atoms.
	The temperatures in (a)-(c) are extracted from the Gaussian part of the fittings.} 
\end{figure}

Figure \ref{fig:1} shows the absorption images taken after 18\,ms of expansion and the corresponding linear density profiles at different final temperatures, showing the phase transition to BECs.
Before the phase transition (a), the atomic distribution was well fitted by a Gaussian function, indicating that atoms in the trap followed the Maxwell-Boltzmann distribution. 
The temperature was estimated to be $679(9)\, \mathrm{nK}$ from the Gaussian width; the number in parentheses indicates standard error.
When the atomic sample was cooled, a dense parabolic peak emerged from a Gaussian background (b), signaling the formation of a BEC.
With bimodal fitting \cite{Davletov2020}, we estimated a temperature of $349(8)\, \mathrm{nK}$ and a BEC fraction of 7(1)\,\%.
The theoretical critical temperature was estimated to be $367 (4)\, \mathrm{nK}$, which is consistent with our observation.
Note that the critical temperature was calculated from BEC theory \cite{Glaum2007}, including the finite size effect, the s-wave interaction effect with a scattering length of $a_s = 110\, a_\mathrm{B}$, which was measured later, and the DDI effect.
In this calculation, we used the measured trap frequencies of $(\nu_x, \nu_y, \nu_z) = (97(1), 226(2), 217(2))\, \mathrm{Hz}$ and an atom number of $1.61(3)\times 10^5$.
Further cooling increased the BEC fraction to 35(1)\,\% (c), and we finally obtained an almost pure BEC (d) containing $5.02(2)\times 10^4$ atoms, where the thermal component was no longer discernible.

The s-wave scattering length for spin-polarized ${}^{151}$Eu was measured via the following method, which was first demonstrated with ${}^{52}$Cr \cite{Griesmaier2006}.
The values to measure are asymptotic velocities of expansion of the condensate with different orientations of the applied magnetic field $\vb*{B}$.
By comparing the ratio of the two asymptotic velocities with numerical simulation, one can estimate $a_s$.
Here, the numerical simulation was based on the Gross-Pitaevskii equation with DDI in the Thomas-Fermi limit \cite{Giovanazzi2006}.
This method has the advantage that the accuracy of the s-wave scattering length is insensitive to that of the atom number or trap frequencies, in contrast to other methods \cite{Mewes1996, PereiraDosSantos2001, Tang2015, Patscheider2022}.

\begin{figure}[htb]
	\includegraphics{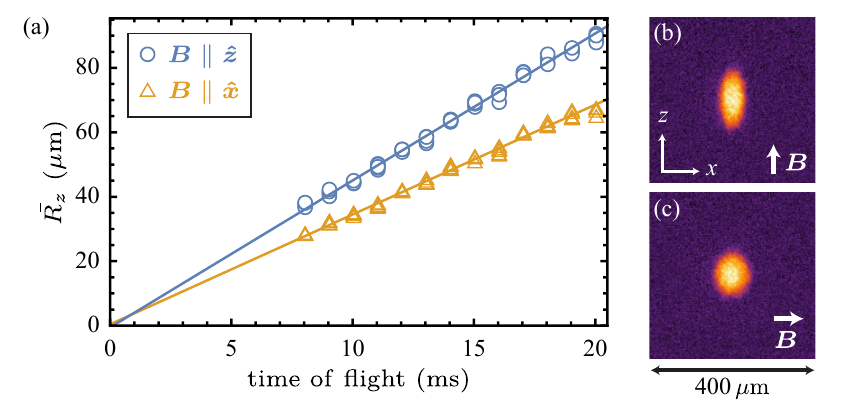}
	\caption{\label{fig:2}
	(a) Measured dependence of the rescaled condensate radius $\bar{R_z}$ on the time of flight.
	The blue open circle (yellow open triangle) is measured $\bar{R_z}$ when the magnetic field is set parallel to the $z$ ($x$) axis.
	The solid line is the linear fit to the data.
	Typical absorption images after 20\,ms-long time of flight are shown in (b) and (c), where the magnetic field is set parallel to the $z$ and $x$ axes, respectively.
	} 
\end{figure}

After creating a BEC, we set a magnetic field in the $x$- or $z$-direction with a magnitude of 0.5\,G.
Then, we turned off the ODTs, let the BEC expand, and maintained the magnetic field for a duration of $5\,\mathrm{ms}$, which was long enough to convert the mean-field energy of the BEC into kinetic energy.
After that, we rotated the magnetic field to the imaging axis and acquired an absorption image at various times.
Figure \ref{fig:2} (a) shows the measured dependence of $\bar{R_z}$ on the time of flight, where $\bar{R_z}$ is the rescaled Thomas-Fermi radius by an atom number \cite{Griesmaier2006}.
Typical absorption images after $20\, \mathrm{ms}$ time of flight are also shown in Fig.~\ref{fig:2} (b) and (c), where the magnetic field is parallel to the $z$ and $x$ axes, respectively.
The effect of DDIs can be clearly seen as the elongation of the BEC in the magnetic field direction \cite{Stuhler2005, Griesmaier2006}.
Upon linear fitting of the experimental data, we obtained the two asymptotic velocities $v_z(\vb*{B}\parallel \vu*{z}) = 4.52(4)\, \mathrm{mm/s}$ and $v_z(\vb*{B}\parallel \vu*{x}) = 3.33(3)\, \mathrm{mm/s}$ and their ratio $v_z(\vb*{B}\parallel \vu*{z}) / v_z(\vb*{B}\parallel \vu*{x}) = 1.357(17)$.
By comparing the ratio with the numerical simulation, we obtained $a_s = 110(4)\, a_\mathrm{B}$ and the corresponding relative DDI strength $\epsilon_{dd} = a_{dd}/a_{s} = 0.54(2)$.

\begin{figure}[htb!]
	\includegraphics{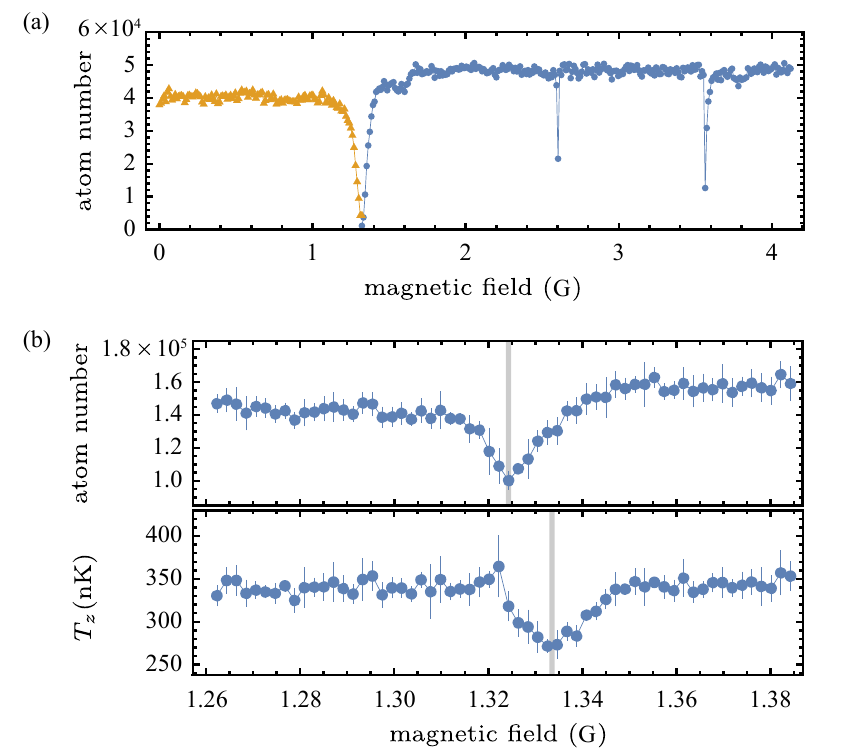}
	\caption{\label{fig:3}
	(a) Feshbach spectrum of the ${}^{151}$Eu BEC with a resolution of 10\,mG.
	To avoid sweeping the magnetic field across the resonance at $B=1.32\,\mathrm{G}$, which would cause unfavorable atom loss \cite{Inouye1998}, data plotted in the blue circle (yellow triangle) were taken by creating a BEC at a higher magnetic field of 3.00\,G (lower magnetic field of 1.00\,G) than the resonance.
	(b) Feshbach spectrum of thermal ${}^{151}$Eu atoms trapped in a horizontal ODT around the $B=1.32\,\mathrm{G}$ resonance.
	The resolution is 2\,mG.
	The measured atom number (top panel) and effective temperature $T_z$ (bottom panel) are plotted as a function of the magnetic field.
	The minimum in the atom number indicates the Feshbach resonance pole, whereas the minimum in the effective temperature $T_z$ indicates zero crossing of the scattering length.
	Both are marked by vertical gray lines.
	} 
\end{figure}

We also performed Feshbach spectroscopy for ${}^{151}$Eu BECs in a low magnetic field of up to 4.1\,G.
After production of the BEC, we changed the magnetic field to a target value, held the BEC for $20\, \mathrm{ms}$ in the trap under the target magnetic field, and measured the residual atom number.
One can find Feshbach resonances as resonant atomic losses.
Figure \ref{fig:3} (a) shows the loss spectrum.
We found one relatively broad resonance ($B=1.32\,\mathrm{G}$) and two narrow resonances ($B=2.60\,\mathrm{G}$ and 3.56\,G).
The width of the broader resonance at 1.32\,G was estimated by Feshbach spectroscopy for thermal ${}^{151}$Eu atoms.
We prepared atoms trapped in the horizontal ODT under a target magnetic field and abruptly reduced the potential depth and spilled out the higher energy atoms.
Since the trap depth decreased more along the $z$-direction than along the other axes due to gravity, anisotropy in the resultant kinetic energy was induced, where the energy along the $z$-axis was the lowest.
We then let the system evolve for $150\, \mathrm{ms}$ to undergo thermalization and measured the residual atom number and effective temperature $T_z$.
Figure \ref{fig:3} (b) shows the loss spectrum and effective temperature $T_z$ as a function of the magnetic field.
The maximal atom loss indicates the Feshbach resonance pole, whereas the minimum in the effective temperature $T_z$ indicates zero crossing of the scattering length, where cross-dimensional thermalization made little progress \cite{Aikawa2012,Kadau2016}.
From the difference between them, we estimated the width of the resonance $\Delta B=10(2)\,\mathrm{mG}$.
We note that the magnetic field was calibrated by spectroscopy on the narrow-line $a\, {}^8\mathrm{S}_{7/2} \leftrightarrow z\, {}^{10}\mathrm{P}_{9/2}$ transition using cold Eu atoms.
The calibration yields a relative uncertainty of $0.4\,\%$, and the absolute error is less than $10\,\mathrm{mG}$ for the magnetic field values around the $B=1.32\,\mathrm{G}$ resonance.

Although we observed Feshbach resonances in the low magnetic field, their origin is different from that of Dy, Er, and Tm \cite{Maier2015, Khlebnikov2019}.
The Feshbach resonances for these atomic species are mainly induced by their orbital anisotropy and exhibit chaotic properties \cite{Maier2015}, whereas Eu has no orbital anisotropy in the ground state.
A theoretical investigation \cite{ZarembaKopczyk2018} predicted that small hyperfine spacings and the coupling between entrance channels (s-wave) and closed channels (d- or g-waves) through DDIs would induce dense Feshbach resonances in a low magnetic field without a chaotic distribution. 
The typical widths of d-wave resonances were estimated to be between $10\, \mathrm{mG}$ and $100\, \mathrm{mG}$, whereas the widths of g-wave resonances were estimated to be below $10\, \mathrm{mG}$.
The spectral widths observed in our experiment were consistent with the theoretical prediction.

\begin{figure}[htb]
	\includegraphics{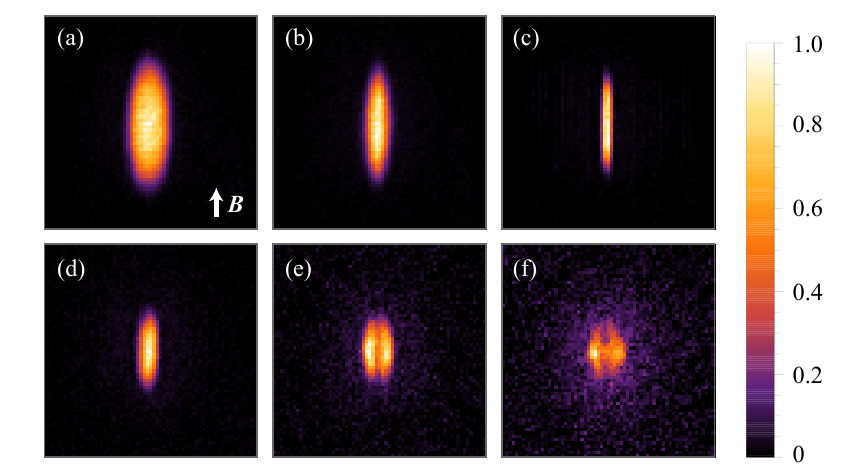}
	\caption{\label{fig:4} 
	Absorption images of Eu BECs in the vicinity of the Feshbach resonance at $B=1.32\, \mathrm{G}$.
	Each of the images is an average of eight pictures.
	The field of view is $200\, \mathrm{\mu m} \times 200\, \mathrm{\mu m}$.
	The color bar shows the optical density scale of the images, where the maximum optical density is set to unity.
	The maximum optical densities are (a) 0.75, (b) 0.86, (c) 0.76, (d) 0.41, (e) 0.15, and (f) 0.10.
	The target magnetic field values of each image are (a) $1.436\, \mathrm{G}$, (b) $1.343\, \mathrm{G}$, (c) $1.331\, \mathrm{G}$, (d) $1.325\, \mathrm{G}$, (e) $1.323\, \mathrm{G}$, and (f) $1.322\, \mathrm{G}$.
	} 
\end{figure}

To demonstrate the control of the scattering length with the Feshbach resonance at $B=1.32\, \mathrm{G}$, we observed the shape of the BEC at a magnetic field slightly higher than the resonance pole.
We first produced pure BECs at a magnetic field of $1.436\, \mathrm{G}$, which is above the Feshbach resonance.
We then linearly ramped down the magnetic field to a variable target value within $20\, \mathrm{ms}$ and turned off the trap.
The magnetic field was kept constant at its target value during the first $15\, \mathrm{ms}$ of the expansion and then set along the probe axis.
After an additional $3\, \mathrm{ms}$ of expansion, we acquired an absorption image of the atomic cloud, which is summarized in Fig. \ref{fig:4}.
Figure \ref{fig:4} (a) shows the absorption image of the BEC at a magnetic field of $B=1.436\, \mathrm{G}$, which is much higher than the resonance.
Upon decreasing the magnetic field, the BEC showed a more anisotropic shape (b) and (c), showing enhancement of the magnetostriction effect, which indicates reduction of the s-wave scattering length \cite{Lahaye2007}.  
Upon further decreasing the magnetic field, the BEC collapsed, and the cloud split into two (d)-(f), as observed in ${}^{52}$Cr BEC \cite{Metz2009}.
Regarding the collapse, a different story was suggested by Jens Hertkorn in the Universität Stuttgart group that these TOF images may be the repulsion of two droplets. A detailed experiment such as in situ observation is required to confirm this possibility.

In conclusion, we have demonstrated a BEC of ${}^{151}$Eu atoms with a large dipole moment $\mu = 7\,\mu_\mathrm{B}$ and a highly symmetric $[\mathrm{Xe}]\ 4f^7 6s^2\ {}^8\mathrm{S}_{7/2}$ electronic ground state.
The scattering length of ${}^{151}$Eu was measured to be $110(4)\, a_\mathrm{B}$.
In addition, we observed a low-magnetic-field Feshbach spectrum and demonstrated control of the s-wave scattering length with resonance.
Since the Feshbach spectrum of Eu is not expected to exhibit chaotic properties, the s-wave scattering lengths for different spin channels can be determined from the spectrum, as in the case of Cr \cite{Werner2005}, which is crucial for quantum simulation of spinor dipolar BECs.
The production of Eu BECs under ultralow magnetic fields will expand the research field of spinor dipolar BEC, which includes the Einstein--de Haas effect \cite{Kawaguchi2006a} and spontaneous circulation in ground-state \cite{Kawaguchi2006}.

\section*{Acknowledgments}
We are grateful to K. Aikawa, J. Hertkorn, and T. Pfau for fruitful discussions.
This work was supported by JST Grant Numbers JPMJPF2015 and JPMJMI17A3 and JSPS Japan KAKENHI Grant Number JP20J21364.


\bibliographystyle{apsrev4-2}
\bibliography{reference}

\end{document}